\documentstyle[twocolumn,prl,aps]{revtex}
\begin{document} 
\title{
Weak Field Phase Diagram for an Integer Quantum Hall Liquid
}
\author{D. Z. Liu$^{1}$, X. C. Xie$^{2}$, Q. Niu$^{3}$}
\address{$^1$
James Franck Institute, University of Chicago, Chicago, IL 60637
}
\address{$^2$
Department of Physics, Oklahoma State University,
Stillwater, OK 74078.
}
\address{$^3$
Department of Physics, University of Texas, Austin, TX 78712
}

\address{\rm (Submitted to Physical Review Letters on 3 April 1995)}
\address{\mbox{ }}
\address{\parbox{14cm}{\rm \mbox{ }\mbox{ }
We study the localization properties in the transition
from a two-dimensional electron gas
at zero magnetic field into an integer quantum Hall (QH) liquid.
By carrying out a direct calculation of the localization length for
a finite size sample using a
transfer matrix technique, we systematically investigate the field and
disorder dependences of the metal-insulator transition in the weak field
QH regime. We obtain a different phase diagram from the one conjectured
in previous theoretical studies. In particular, we find that:
(1) the extended state energy $E_{c}$
for each Landau level (LL)
is {\it always} linear in magnetic field;
(2) for a given Landau level and disorder configuration
there exists a critical magnetic
field $B_{c}$ below which the extended state disappears;
(3) the lower LLs are more robust to the metal-insulator
transition with smaller $B_{c}$. We attribute the above results
to strong LL coupling effect.
Experimental implications of our work are discussed.
}}
\address{\mbox{ }}
\address{\parbox{14cm}{\rm PACS numbers: 71.30.+h, 73.20.Jc, 73.40.Hm}}
\maketitle

\makeatletter
\global\@specialpagefalse
\def\@oddhead{REV\TeX{} 3.0\hfill Xie Group Preprint, 1995}
\let\@evenhead\@oddhead
\makeatother

It is very important to understand the localization properties in the
transition from two-dimensional
electron gas (at zero magnetic field) into an integer quantum Hall liquid
\cite{prange:qhe}.
According to the scaling theory of localization \cite{anderson:scal}
all electrons in a two-dimensional system are localized in the absence of
magnetic field. When the two-dimensional electron system is subject to
a strong perpendicular magnetic field, the energy spectrum becomes
a series of impurity broadened Landau levels.
Extended state appears in the center of each Landau band,
while states at other energies are localized.
This gives rise to the integer
quantum Hall effect.
The interesting issue is to understand the evolution of the extended states
in the weak field regime as the magnetic field goes to zero where all extended
states disappear.

There could be two scenarios for the fate of the extended states as
$B\rightarrow 0$. The first one was proposed by Kivelson, Lee, and Zhang
\cite{klz} in their global phase diagram of the quantum Hall effect.
According to this phase diagram,
in a strongly disordered quantum Hall system, the extended
states stay with the center of the Landau bands at strong magnetic field,
but float up in energy at small magnetic field and go to infinity as
$B\rightarrow 0$. This phase diagram is consistent with the
semiclassical argument put forth by  Khmelnitskii \cite{russian}
and Laughlin \cite{laughlin}.

In this letter, we are proposing an alternative scenario for the behavior of
the extended states at weak magnetic field limit. In our picture, each
extended state is simply destroyed by strong disorder at a critical magnetic
field instead of floating up in energy.
By carrying out a direct calculation of the localization length for
a finite size sample using a
transfer matrix technique, we systematically investigate the field and
disorder dependence of the metal-insulator transition in the weak field
quantum Hall regime.
We find that:
(1) the extended state energy $E_{c}$
for each Landau level (LL)
is {\it always} linear in magnetic field;
(2) for a given Landau level and disorder configuration
there exists a critical magnetic
field $B_{c}$ below which the extended state disappears;
(3) the lower LLs are more robust to the metal-insulator
transition with smaller $B_{c}$. We attribute the above results
to strong LL coupling effect.

\begin{figure}
 \vbox to 6.0cm {\vss\hbox to 8cm
 {\hss\
   {\includegraphics{/home/ldz/tek/paper/psd02/f1.ps}
   }
  \hss}
 }
\caption{
Sketch of our proposed weak field phase diagram for an integer quantum
Hall liquid, where $E$ is the Fermi energy, and $B$ the magnetic field.
Shaded area represents the metallic regime.
\label{wkb-f1}}
\end{figure}

Our results can be summarized as the phase diagram presented in
Fig. \ref{wkb-f1}. At strong magnetic field, the extended states appear in the
center of Landau bands ($E_n=(n+1/2)\hbar\omega_c$, where $\omega_c=eB/mc$).
As the magnetic field decreases, the energies for the extended states also
decrease. In the weak magnetic field regime, the extended states disappear
when the magnetic field is less than a critical magnetic field which depend on
the parameters of the quantum Hall system (such as sample size and disorder
strength). The higher the Landau level is, the larger the critical magnetic
field for destroying the extended states. The extended state associated with
the lowest Landau band is the most robust and survives at very low magnetic
field and is
finally destroyed before the magnetic field reaches zero. In other
words, in our scenario, the energies for the
extended states never ``float'' up and there is a critical magnetic field to
create any extended states for a given disorder configuration.

There have been a number of experimental
attempts \cite{jiang-1,wang,kravechenko,jiang-2,sergey}
to address the transition
of delocalized states at the weak magnetic field limit.
Earlier experiments \cite{jiang-1,wang} on strongly disordered two-dimensional
electron gas (2DEG) have demonstrated a transition
from an Anderson insulator at $B=0$ (with all states localized) to quantum
Hall conductor at strong magnetic field. These experiments serve as direct
evidence for magnetic field induced delocalization in 2D systems. They are
certainly consistent with our weak field phase diagram.

Two very recent experiments on low mobility gated GaAs/AlGaAs
hetereostructure\cite{jiang-2} and high mobility Si samples\cite{sergey} have
reported floating up of the delocalized states in the carrier-density --
magnetic-field plane. The authors in Ref.\cite{jiang-2}
claim that their experimental result
``unambiguously'' demonstrated that the ``energy'' of the delocalized state
(extended states)
floats up as $B\rightarrow 0$. However, as we will show
in the latter part of this paper, floating up of the carrier density does not
necessarily imply the same behavior of the energy. Strong disorder scattering
and Landau level mixing effects (which are important at the weak field limit)
could contribute to anomalous behavior of the
carrier density while keeping the extended states at the center of
lowest Landau band.
In fact, we will demonstrate that these experimental results
\cite{jiang-1,wang,kravechenko,jiang-2,sergey}  are all
consistent with our weak field phase diagram (Fig. \ref{wkb-f1})
for the integer quantum Hall liquid.

In the following, we briefly outline our model and technique to calculate the
localization length.
We model our two-dimensional system in a very long strip geometry
with a finite width ($M$) square lattice
with nearest neighbor hopping. Periodic boundary condition in the width
direction is used to
get rid of the edge extended states. The disorder potential is modeled by
the on-site white-noise potential $V_{im}$ ($i$ denotes the column
index, $m$ denotes the chain index) ranging from $-W/2$ to $W/2$.
The effect magnetic field appears in the complex phase of the hopping term.
The strength of the magnetic field is characterized by the flux per plaquette
($\phi$)
in unit of magnetic flux quanta ($\phi_o=hc/e$).
The Hamiltonian of this system can be written as:
\begin{eqnarray}
{\cal H} &= &\sum_i\sum_{m=1}^{M} V_{im}|im><im| \\
& &+\sum_{<im;jn>}\left[
t_{im;jn}|im><jn|+t^{\dagger}_{im;jn}|jn><im|\right],\nonumber
\end{eqnarray}
where $<im;jn>$ indicates nearest neighbors on the lattice. The
amplitude of the hopping term is chosen as the unit of the energy.
For a specific energy $E$, a transfer matrix $T_{i}$ can be easily set up
mapping the wavefunction amplitudes at column ${i-1}$ and $i$ to those
at column $i+1$, {\it i.e.}
\begin{equation}
\left( \begin{array}{c} \psi_{i+1} \\ \psi_{i} \end{array} \right) =
T_{i}
\left( \begin{array}{c} \psi_{i} \\ \psi_{i-1} \end{array} \right) =
\left( \begin{array}{cc} H_{i}-E & -I \\ I & 0 \end{array} \right)
\left( \begin{array}{c} \psi_{i} \\ \psi_{i-1} \end{array} \right) ,
\end{equation}
where $H_i$ is the Hamiltonian for the $i$th column, $I$ is a $M\times
M$ unit matrix.
Using a standard iteration algorithm \cite{ldz:bloc}, we can
calculate the Lyapunov exponents for the transfer matrix $T_{i}$.
 The localization length $\lambda_M(E)$ for energy
$E$ at finite width $M$ is then given by the inverse of the smallest
Lyapunov
exponent. In our numerical calculation, we choose the sample length to
be over $10^4$ so that the self-averaging effect automatically takes
care of the ensemble statistical fluctuations.

\begin{figure}
\vspace{2mm}
 \vbox to 5.0cm {\vss\hbox to 6.5cm
 {\hss\
   {\includegraphics{/home/ldz/tek/paper/psd02/f2a.ps}
   }
  \hss}
 }
\vspace{2mm}
 \vbox to 5.0cm {\vss\hbox to 6.5cm
 {\hss\
   {\includegraphics{/home/ldz/tek/paper/psd02/f2b.ps}
   }
  \hss}
 }
\caption{
Finite-size localization length ($\lambda_M$) at different weak magnetic
field (from the lowest curve to the highest one, (a)
$\phi = 1/13, 1/9, 1/7, 1/5$; (b) $\phi= 1/33, 1/31, 1/29, 1/27, 1/25$)
 for fixed disorder strength $W=1$.
The localization length for
higher magnetic field is scaled by a factor proportional to the magnetic field
from the one for next smaller field to show the linear dependence in
magnetic field.
\label{wkb-f2}}
\end{figure}

In Fig. \ref{wkb-f2}, we present the localization length at various weak
magnetic fields for a finite size sample ($M=32$). Due to the symmetry of the
lattice model, only the lower energy branch results are shown here. In the
absence of disorder, the energy band for a tight-binding lattice should break
up into $q$ subbands in a magnetic field with $\phi =1/q$ (where $q$ is an
integer). In a continuous model, the subbands in the lower energy branch
correspond to Landau levels.
In the presence of disorder, for a continuous model, each Landau level evolves
into impurity broadened Landau band with extended state in the center of each
Landau band.
We have similar effects in the lattice model as presented in
Fig. \ref{wkb-f2}. The maximas in the finite-size
localization length are the locations of the extended states.
In Fig. \ref{wkb-f2}, we can clearly see
that the energy band does break up into small subbands (Landau bands)
at corresponding
magnetic field, {\it i.e.} at $\phi =1/q$ there are $q$ subbands.
The extended states appear at the centers of these subbands
and their energies (count from the band edge $E=-4.0$)
are linear in $B \propto \phi$,
resembling the behavior of a continuous model.

\begin{figure}
 \vspace{2mm}
 \vbox to 5.0cm {\vss\hbox to 6.5cm
 {\hss\
   {\includegraphics{/home/ldz/tek/paper/psd02/f3.ps}
   }
  \hss}
 }
\caption{
Finite-size localization length ($\lambda_M$) for different disorder strength
at magnetic field $\phi =1/11$ and  $W=1, 2, 3, 4, 5$.
(from the highest curve to the lowest one).
The localization length for weaker disorder strength is scaled by
a factor of 500 from the one for next stronger disorder strength.
\label{wkb-f3}}
\end{figure}

We now address the effect of disorder strength on the extended states at the
weak magnetic field limit. As presented in Fig. \ref{wkb-f3}, the localization
length decreases as strength of disorder increases. In a fixed magnetic field,
the extended states in higher energy subbands (Landau bands) are destroyed in
the presence of strong enough disorder, while at the same time
 the width of the extended-state
bands in the lower energy subbands (Landau bands) become narrower. The
extended states in the lowest energy subband (Landau band) are the most robust
and nevertheless destroyed at very strong disorder strength.
In the entire range of the disorder strength shown in Fig. \ref{wkb-f3},
 even though the localization
lengths are changed by several order of magnitude, the energies of the extended
states still stay with corresponding energy band without floating up at strong
disorder limit.
We can conclude that in a continuous model with fixed disorder strength,
the extended states always stay in
their corresponding Landau bands, and are finally destroyed below certain
critical magnetic field (the higher the Landau level is, the larger the
critical field). This is the basis for our proposed weak field phase diagram
(Fig. \ref{wkb-f1}).
We believe that the localization transition here is caused by Landau level
coupling effect which is more severe at weak field. For our model in Eq.(1),
the zero-field level broadening is\cite{economou}
\begin{equation}
 \Gamma = {W^{2} \over {6\pi E}} K({4t\over {E}}),
\end{equation}
where $t$ is hopping amplitude which is set to unity and $K(x)$ is the
complete elliptical integral of first kind.
The peaks in Fig.\ref{wkb-f3}
start to disappear when $\Gamma \simeq \omega _{c}$, {\it i.e.}
when the Landau levels start to couple together.
We should mention that the global phase diagram\cite{klz}
is based on the level floating up theory. However, that theory
has a crucial assumption:  the
extended states
are concentrated at discrete levels.
This assumption is certainly true for strong
field and smooth random potentials.  To our knowledge, its validity for
short range potentials has not been firmly established.
The situation for weak field and strong disorder is totally unclear.
There exists
no theory showing that the extended states still have to remain at discrete
levels when the Landau levels are strongly mixing.

In the experiments observing transition from Anderson localization to quantum
Hall conductor \cite{jiang-1,wang}, only the lowest Landau level plateau was
observed ($\nu =2$ for spin unresolved 2DEG) which is consistent with our
argument that the delocalization states in the lowest Landau band are the most
robust. We propose similar experiment in strong magnetic field to test our
scenario that higher Landau level QH plateau should emerge in sequence.

To explain the anomalous floating up of the carrier density in the recent
experiment by Glozman {\it et al} \cite{jiang-2}, we should include the Landau
level mixing effect at small magnetic field in the presence of strong disorder
scattering.  In the weak disorder limit, the Landau level broadening is much
smaller comparing with the inter-Landau level energy difference as shown by
Ando {\it et al} \cite{ando-dos} in a simple Born-approximation calculation
without any inter-Landau level coupling.
However in a realistic situation, the disorder potential is Coulomb long range
type and Landau levels are much more broadened than in the short-range impurity
case, and therefore, the effect of  Landau level mixing is much more important
as demonstrated by Xie {\it et al} \cite{xie-dos}. In another word, the
localized tail
of higher Landau bands could well extended into the lower Landau bands.
So the electrons have to fill up the tail of higher Landau bands before
reaching the extended states in the center of the lowest Landau band, and
therefore,  the
real filling factor could be greater than 2 (for spin unresolved 2DEG)
when the QH plateau for the lowest Landau level is observed.
The critical point where the carrier density starts to float up is the moment
where inter-Landau level mixing shows up.
The experimental fact that no floating was observed in higher mobility samples
\cite{jiang-2} strongly demonstrate the essential role of the
Landau level mixing in the floating up of the carrier density.
In the following we will demonstrate the above argument at a more quantitative
level. Let us first consider the high field limit such that Landau levels
are well separated. In this limit, the electron density $\rho _{c}$
below $E_{c}$ (energy
of the extended state for the lowest Landau level) is proportional
to the Landau level degeneracy. Thus, $\rho _{c}$ decreases linearly
with decreasing $B$.
At the weak field limit with strong Landau level coupling,
the experimental density of states in {\it GaAs} samples can be well
described by the Lorentzian form\cite{ashoori}
\begin{equation}
g(E)={1 \over{2 \pi l^{2}}} \sum_{n}{\Gamma _{n}\over
{(E-E_{n})^{2}+\Gamma_{n}^{2}}}
\end{equation}
and
\begin{equation}
\rho _{c}=\int _{-\infty}^{E_{0}} g(E) dE \propto \omega _{c}\sum_{n}(1-{2\over
{\pi}}arctan{n\omega _{c}\over {\Gamma_{n}}})
\end{equation}
where $l$ is the magnetic length, $\omega _c$ is the cyclotron energy
and $E_{n}=(n+1/2)\omega_{c}$ is the energy for the $n$th Landau level.
$\Gamma _{n}$ is the level broadening for $n$th Landau level
which is found experimentally\cite{ashoori} to be independent of
magnetic field.
As $B$ (or $\omega _{c}$) decreases with fixed $\Gamma _{n}$,
more terms will be contained in the summation which
makes $\rho _{c}$ to increase.
In the limit of
infinite summation over Landau levels, $\rho _{c}$ diverges.
This simple model calculation shows that
in the high field limit $\rho _{c}$ goes down linear
with decreasing $B$, and in the zero field limit it approaches infinity.
Therefore, $\rho _{c}$ has to {\it float up} at a certain magnetic field.
We should mention that more realistic calculations\cite{xie-dos} are needed
in order to make quantitative comparisons with the experimental results.

In conclusion, we have proposed an alternative phase diagram for the integer
quantum Hall system in the weak field limit. We have demonstrated that there
exists a critical magnetic field to delocalize any two-dimensional disordered
system. The extended states always stays with their corresponding Landau bands
with those in the Lowest Landau bands the most robust. In the strong disorder
limit, Landau level mixing effect could contribute to the floating up of the
carrier density even though the energies of the extended states never float up.

D.Z. Liu is supported by NSF-DMR-94-16926 through the Science and Technology
Center of Superconductivity.


\end{document}